\def\cm2{cm$^{-2}$}

\def\c2{C~{\sc ii}}
\def\c4{C~{\sc iv}}
\def\fe2{Fe~{\sc ii}}
\def\fe3{Fe~{\sc iii}}
\def\mg1{Mg~{\sc i}}
\def\mg2{Mg~{\sc ii}}
\def\si2{Si~{\sc ii}}
\def\si4{Si~{\sc iv}}
\def\al2{Al~{\sc ii}}
\def\al3{Al~{\sc iii}}
\def\o1{O~{\sc i}}
\def\n1{N~{\sc i}}
\def\h1{H~{\sc i}}

\def\approxlt{\mathrel{\spose{\lower 3pt\hbox{$\sim$}}
        \raise 2.0pt\hbox{$<$}}}
\def\approxgt{\mathrel{\spose{\lower 3pt\hbox{$\sim$}}
        \raise 2.0pt\hbox{$>$}}}

\documentclass[article]{has40}
\usepackage{graphicx}
\usepackage{amssymb}
\tabletypesize{\normalsize}  

\def\plotone#1{\centering \leavevmode
\includegraphics[width=.95\columnwidth]{#1}}

\def\plotone#1{\centering \leavevmode
\includegraphics[width=.95\columnwidth]{#1}}

\shortauthors{Molnar et al.}
\shorttitle{Fast Contact Binary Stars}

\begin{document}
\large    
\pagenumbering{arabic}
\setcounter{page}{55}

\title{The Short Period End of the Distribution of Contact Binary Stars}

%
%
\author{{\noindent Lawrence A. Molnar{$^{\rm 1}$}, Daniel M. Van Noord{$^{\rm 1}$}, and Steven D. Steenwyk{$^{\rm 1}$}\\
\\
{\it (1) Department of Physics and Astronomy, Calvin College, Grand Rapids, MI, USA} 
}}
%
%
\email{(1) lmolnar@calvin.edu}

\begin{abstract}
Contact binary systems (also known as W UMa systems) consist of a pair of hydrogen-burning dwarf stars orbiting each other so closely that they share a common envelope.  Although they are relatively common, there is as yet no established consensus on the principle evolutionary questions surrounding them: how do they form, how do they evolve over time, what do they become?

One observational clue to their evolutionary history has been the abrupt termination of the orbital period distribution around 5.2 hours.  We have undertaken an observational study of this by 1) discovery of fast W UMa systems in our Calvin-Rehoboth Observatory data archive, 2) follow-up with the Calvin-Rehoboth Observatory of candidate fast systems from the Catalina Sky Survey, and 3) follow-up of other reports of potentially fast systems in other recently published surveys.  We find the follow-up to have been particularly important as many surveys taken for other purposes lead to ambiguous or incorrect claims for periods less than five hours.

Our results to date may be characterized as showing two distinct components: the steeply decaying tail associated with the previously known cutoff along with a low-amplitude, but apparently uniform distribution that extends down to 3.6 hours.  The confirmation at greater sensitivity of the abruptness of the cutoff seems to imply that the dominant mechanism for system formation (or the mechanism that determines system lifetime) does have a strong period dependence.  At the same time, there appears to be a second mechanism at work as well which leads to the formation of the ultrafast component of the histogram.

\end{abstract}

\section{Introduction}
Although contact binary systems are relatively common, there is as yet no established consensus on the principle evolutionary questions surrounding them: how do they form, how do they evolve over time, what do they become (see for example Eker et al. 2008 and reference therein)?
The observed distribution of orbital periods provides a test for any model.
In particular, the apparently sharp cutoff on the short-period side around 5.30 hours (the value for CC Com) has attracted the attention of modelers (see for example Stepien 2006).
Recent surveys have pushed the record to 5.23 hours for field stars (Rucinski 2007) and 5.17 for cluster stars (Weldrake et al. 2004).
An early find in our own survey of variable stars in an archive of asteroid rotation observations (described by Van Noord et al. in these proceedings) pushed the limit still further.  VSX J205508.0+082951 (discovered by Calvin student Andrew Hess) has an orbital period of 5.06 hours (see its light curve in Figure 1).
This latter find led to our initiating a study of this cutoff.
In particular, we noted the need to be sure observations were sufficient to preclude the  possibility that the cutoff was due to an observational bias against shorter period (and hence fainter) systems.
Likewise, we noted the need to be careful interpreting observations from surveys not optimized for discovery of fast, red systems.

In the following sections we describe the theoretical question more fully, our efforts to search for fast systems both in our archive and in others, and the implications of the resulting frequency histogram.

\begin{figure*}
\centering
\plotone{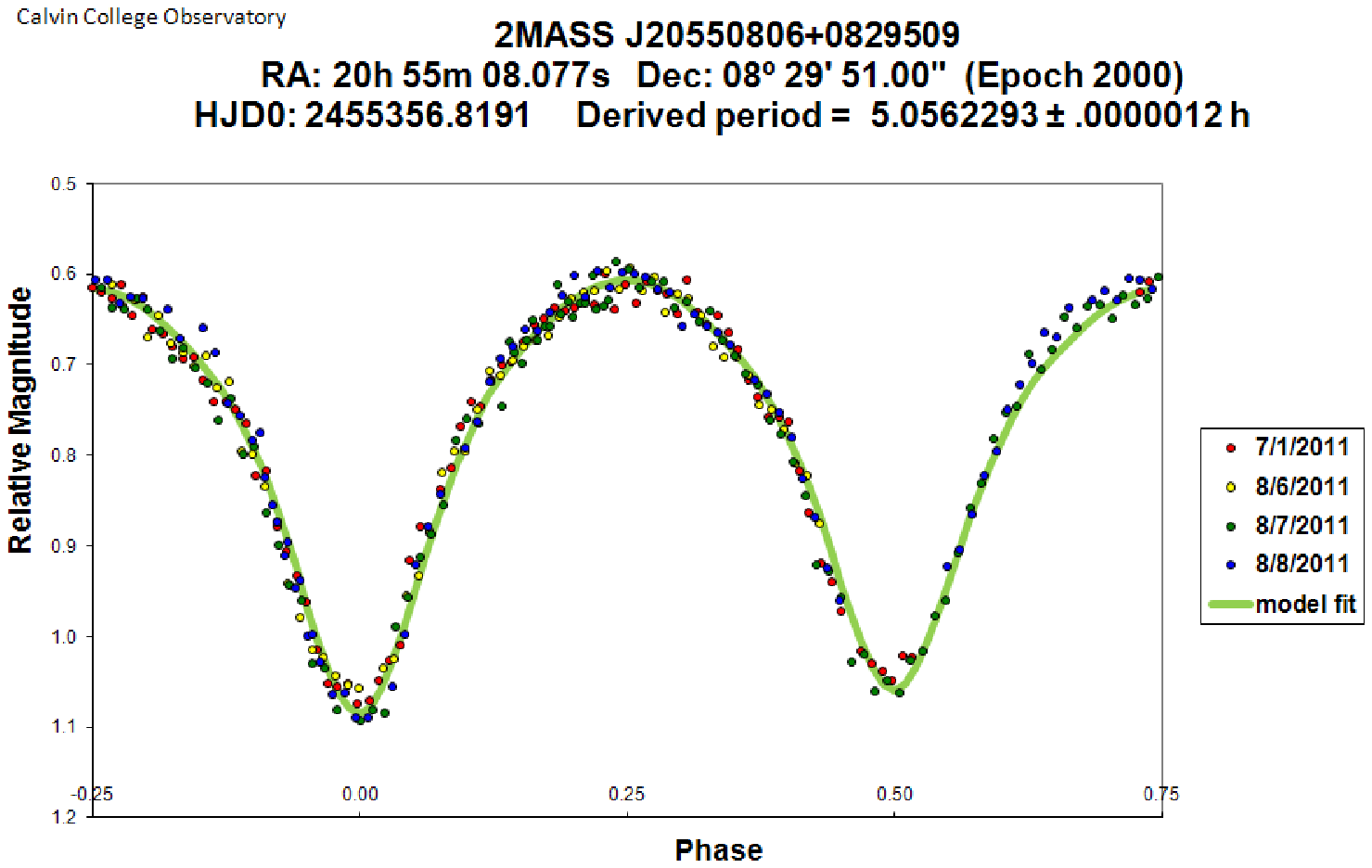}
\vskip0pt
\caption{The light curve of VSX J205508.0+082951.  With an orbital period of 5.06 hours, it was the fastest contact system in the Variable Star Index in 2011.}
\label{o1039a}
\end{figure*}

\section{The evolutionary status of contact binary stars}
The period distribution of contact binary stars in the Variable Star Index\footnote[2]{www.aavso.org/vsx/} (VSX) as of 2012 is shown in Figure 2.
Given the narrow range of stellar radii for which binary systems of a given orbital period can be in contact, the spectral type of the primary star is necessarily well correlated with the orbital period in these systems.
This correlation can be found in the observational catalog of Csizmadia and Klagyivik (2004).
Furthermore, thermal contact forces the companion stars to have surface temperatures very close to those of the primaries.
Hence the system luminosity is necessarily well correlated with the orbital period as well.
This leaves open the possibility that the decreasing frequency on the left side of Figure 2 is simply an observational bias against short period systems.

\begin{figure*}
\centering
\plotone{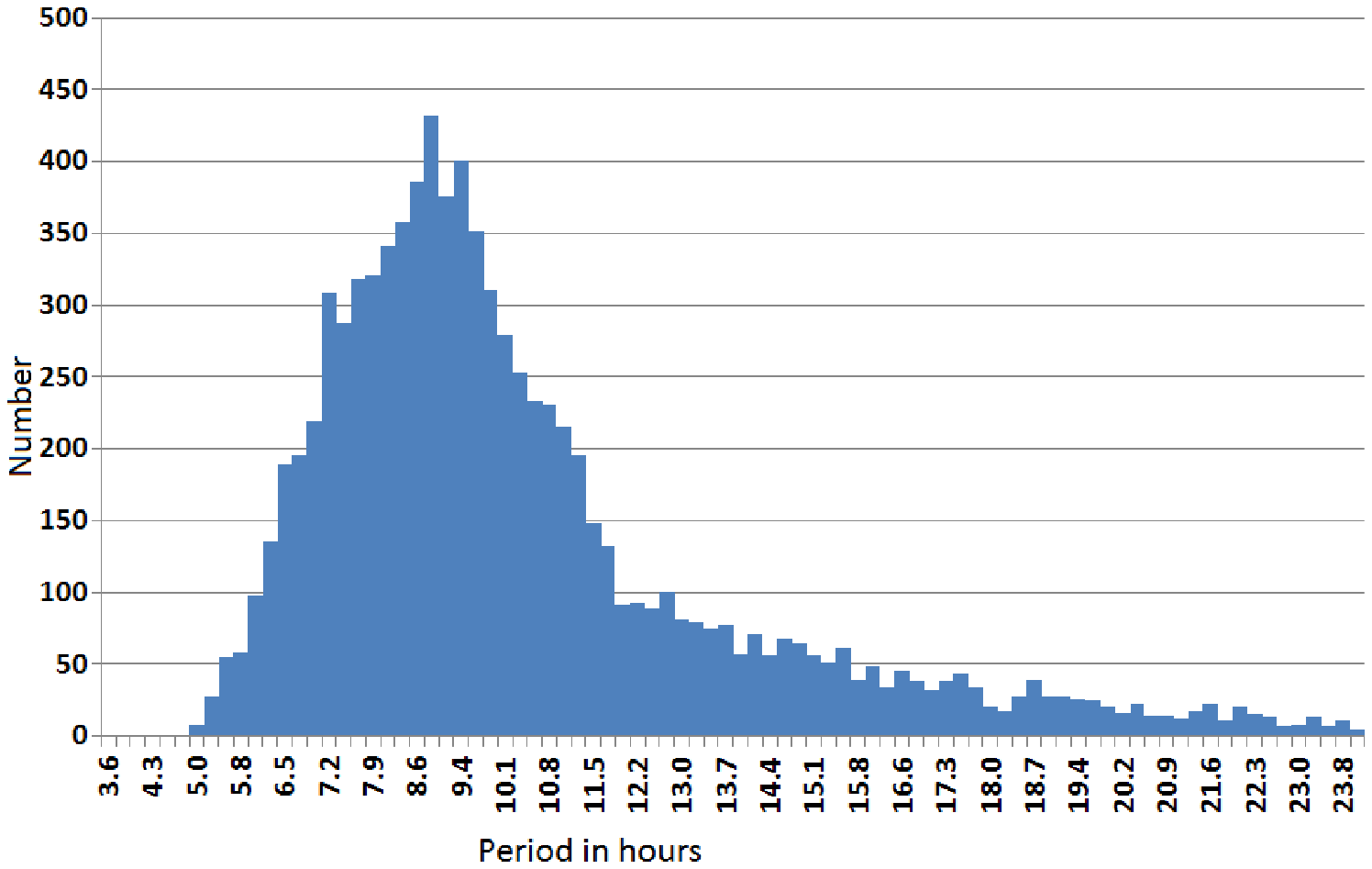}
\vskip0pt
\caption{The period distribution of contact binary stars in the Variable Star Index.  The sharp cut off on the left is near five hours.  The period range up to this cutoff is the subject of this work.}
\label{o1039b}
\end{figure*}

Beyond observational effects, there are three physical mechanisms that can be reflected in the period distribution: system evolution, lifetime, and formation.  We briefly consider each in turn.
By evolution, we consider an analogy with compact binary systems.
With angular momentum loss, orbital periods generally diminish over time.
We can look for such a trend by considering the correlation of mass ratio with orbital period.
If, for example, the companion star loses mass to the primary over time, this could result in a positive correlation.
However, we find no significant correlation in the data from the Csizmadia and Klagyivik (2004) catalog.
Theoretically, there is little reason to expect a correlation, as the requirement of contact limits the range of period change.

By lifetime, we mean that short-lived systems will be less numerous than others just as is found in a Hertzsprung-Russell diagram.
However, the stellar lifetimes of systems with orbital periods less than five hours (systems with late K and M primaries) are longer than those of the more numerous systems.
Certainly, there is no sudden change at five hours to account for a very sharp feature.

Finally, by formation we refer to the likelihood of forming systems of various periods in the first place.
Stepien (2006) suggested that contact binaries evolve from detached systems via angular momentum loss, and that a sharp cutoff can result from the increasing timescale of loss for those systems.
One test of this model would be to look carefully at the period distribution of noncontact systems.
The low mass systems that have not yet had time to come into contact should be seen as an overabundance of noncontact systems.
There is as yet no observational support for this.

\section{Observational search for fast contact binary systems}
We define our search as specifically a search for contact binaries faster than CC Com.
Ultimately, a study of the entire period distribution is of great interest, but will require careful consideration of observational selection biases.
By studying just the region near the cutoff, we can draw conclusions without consideration of observational biases as they do not depend on period strongly enough to create a sharp cutoff.

The first data set we searched was our own.
The data and search methods are described in these proceedings (Van Noord et al.).
The strength of this data set is the depth (17 to 19 magnitude) and the cadence (complete coverage in a single night of nearly one full orbit at a minimum).
The disadvantage is the limited sky coverage (approximately 50 square degrees).
One additional fast object was found (by Calvin student Rick McWhirter in 2012): VSX J061850.4+220511 (Figure 3).
By combining three nights of observing with the Calvin 0.4 m in New Mexico with the sparser Catalina Sky Survery (CSS) data (Drake et al. 2012), a period was established to a part in two million.

\begin{figure*}
\centering
\plotone{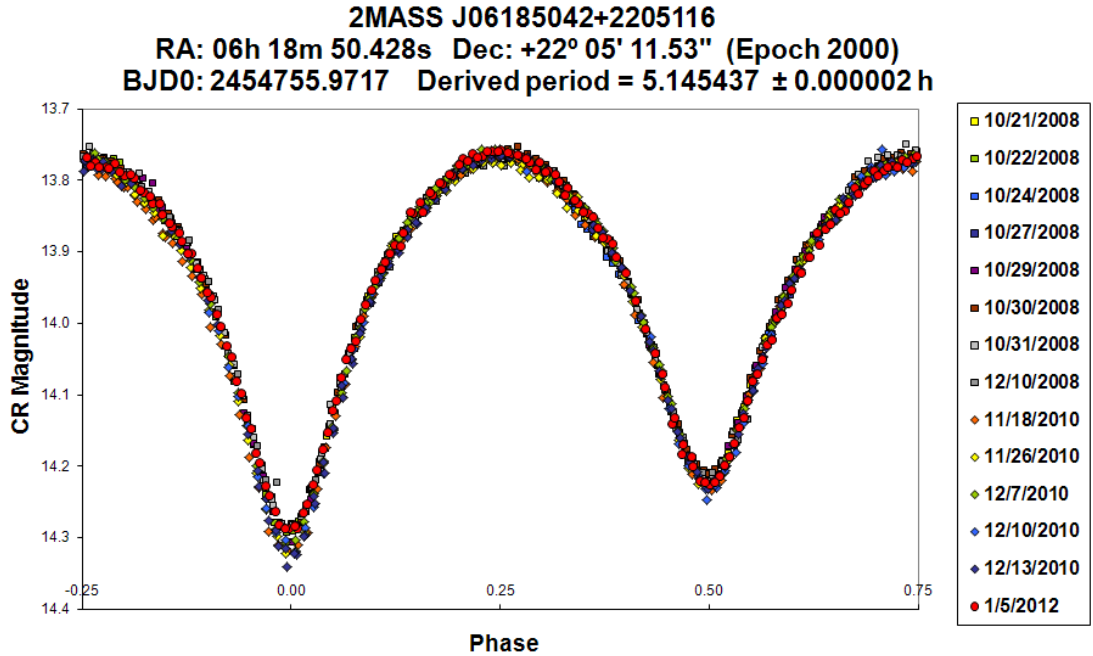}
\vskip0pt
\caption{The light curve of VSX J061850.4+220511.  With an orbital period of 5.15 hours, it is the second fastest system found in the Calvin asteroid rotation archive.}
\label{o1039c}
\end{figure*}

The second data set we searched was the Catalina Sky Survey (Drake et al. 2009).
Andrew Drake did a preliminary period analysis of this much larger catalog (although with much sparser data on most stars).
We systematically followed up his list of potentially fast systems with intense short spacing observations to determine the period, the shape of the light curve, and the color of each system.
A paper with a complete description of this work is in preparation.
Here we note that only 12 of the 34 systems followed up were found to have the proposed period and to be contact systems.
Based on color and light curve shape, a number were classified as pulsating ($\delta$ Scuti) stars.
Others were found to have periods longer or shorter than the proposed values.
Figure 4 shows the light curve of the fastest contact system in the list (and the fastest confirmed system we are aware of today): V0811+3119.
Unlike the previous incremental changes in the period limit, this system's period of 3.75 hours is much faster than any previous period.

\begin{figure*}
\centering
\plotone{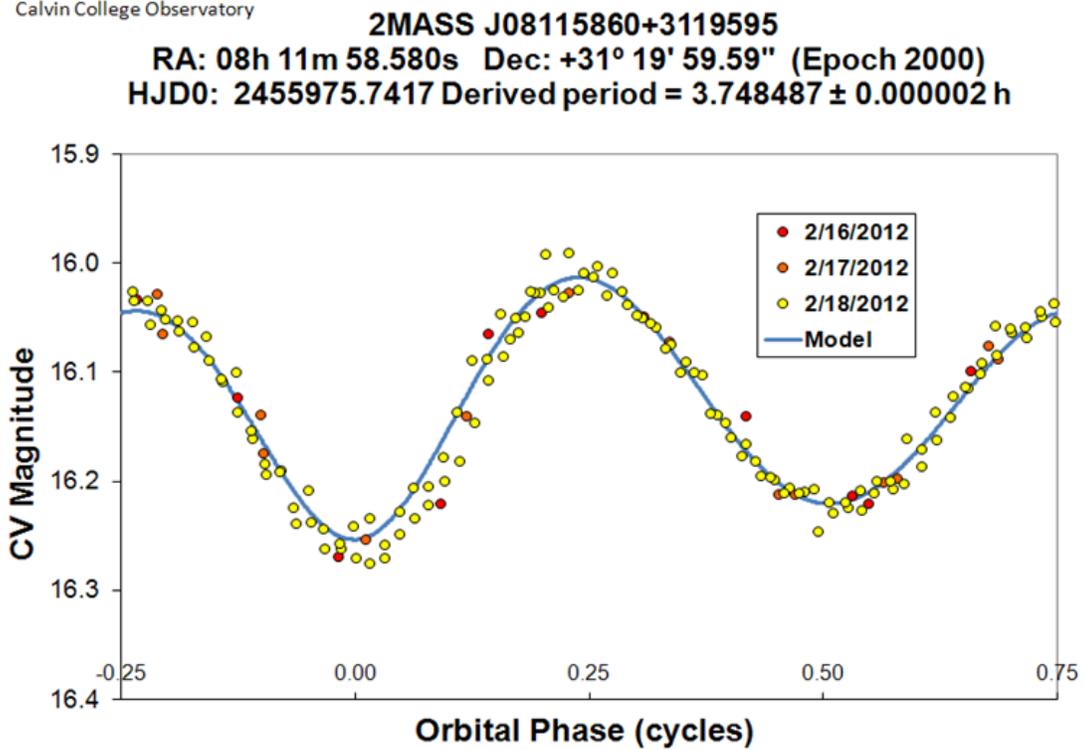}
\vskip0pt
\caption{The light curve of V0811+3119.  With an orbital period of 3.75 hours, it is the fastest confirmed contact system, much faster than CC Com.}
\label{o1039d}
\end{figure*}

Finally, we followed up other systems that were reported as fast contact binary systems in the VSX.
These came from the Kepler, SDSS, and NSVS surveys.
Follow up consisted of consideration of 2MASS colors and CSS photometry where available, or additional observations of our own as needed.
For the Kepler survey, the data quality and cadence were ideal, so the light curve shapes and periods were excellent.
However, none of the 24 systems reported as fast contact systems were binaries.
The light curve shapes, particularly the very low amplitudes (usually fractions of a percent) and blue colors pointed towards pulsating systems.
We attribute these misidentifications to an automated approach to identification rather than to any limitation of the data.
Only one of eleven NSVS candidates turned out to be a fast contact system.
Limitations in data cadence and quality were the important factors here.
Finally, one in seven SDSS candidates turned out to be a fast contact system.
Limitations in data cadence made period determination difficult, and color information was neglected that pointed towards hot, pulsating stars.

In Figure 5, we present the period distribution of confirmed fast contact binaries (again defined as systems faster than CC Com).
We interpret the 12 systems in the rightmost bin as the tail of the sharp period cutoff.
The remaining six systems seem to be an unrelated uniform component seen here for the first time.

\begin{figure*}
\centering
\plotone{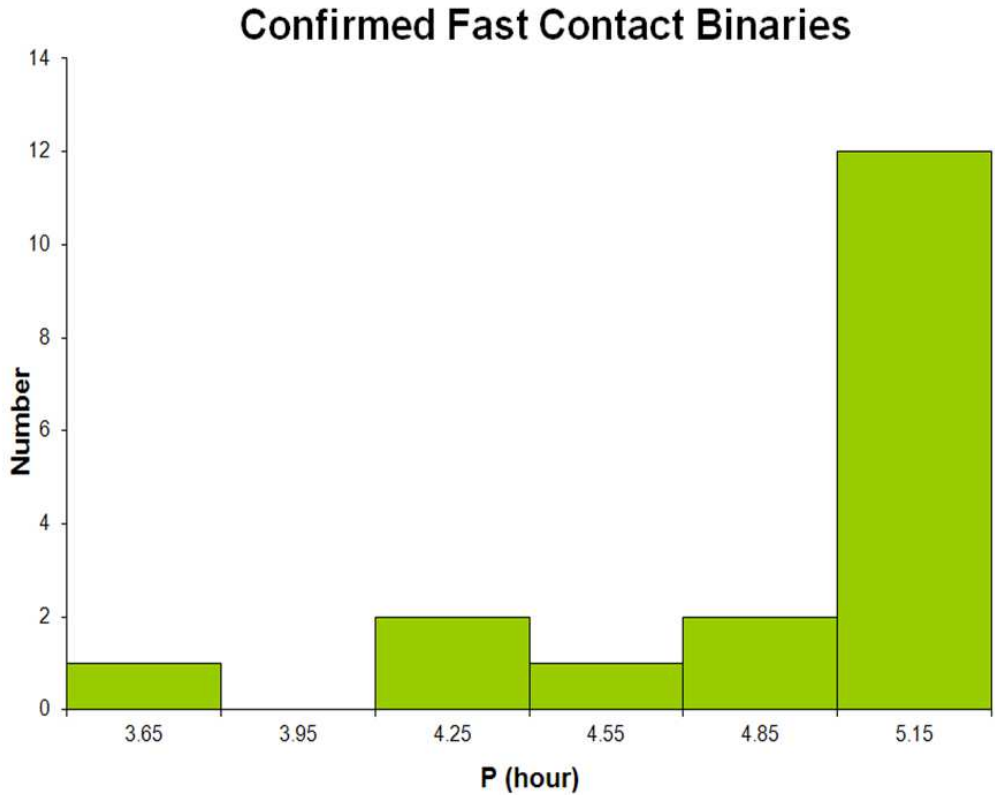}
\vskip0pt
\caption{The period distribution of confirmed fast contact binary star systems.  We interpret the 12 systems in the rightmost bin as the tail of the sharp period cutoff.  The remaining six systems seem to be an unrelated uniform component seen here for the first time.}
\label{o1039e}
\end{figure*}

\section{Discussion and conclusions}
The first conclusion of this work is that the low percentage of confirmations underlines the importance of followup of candidate systems found with sparse data by observing with short time spacings.

We describe the six systems below five hours as a new, uniform component.
Had they been simply a tail of the five hour cutoff, they would have dropped off more gradually.
This only strengthens the previous assertion that the cutoff (which is now revised from 5.2 down to 5.0 hours) is very sharp.

We have no immediate physical understanding of either feature.
The limitations of evolutionary, lifetime, and formational explanations given at the outset remain.
Given the liberty to speculate, though, we suggest theoretical exploration of ``formation in place."
That is to say, perhaps contact binaries are not gradually driven together by angular momentum loss, but rather are rapidly driven together at the time of formation, much as massive planets are now routinely found to have done.
Bilir et al. (2005) give evidence that at least some contact binaries form relatively early.
This new scenario may account for the small number of noncontact systems observed.
Finally, it points to a study of formation dynamics as a place to seek the keys to the period distribution shape.

\section{Acknowledgments}
We thank Andrew Drake for sending us candidate systems from the CSS to follow up observationally.
DMV and SDS acknowledge support from the Michigan Space Grant Consortium.  DMV has also received support from the Calvin College Integrated Science Research Institute.

\end{document}